\documentclass[10pt,twocolumn]{IEEEtran} 

\usepackage{graphicx}
\usepackage{amssymb}
\usepackage{amsmath}
\usepackage{amssymb}
\usepackage{array}
\usepackage{xy}
\usepackage{amsmath}
\usepackage{amsfonts}
\usepackage{latexsym}
\usepackage{graphicx}

\def\BibTeX{{\rm B\kern-.05em{\sc i\kern-.025em b}\kern-.08em
    T\kern-.1667em\lower.7ex\hbox{E}\kern-.125emX}}

\newtheorem{thm}{Theorem}[section]

\newtheorem{lem}[thm]{Lemma}

\begin{document}

\title{Joint Network-Source Coding: An Achievable
Region with Diversity Routing}
\author{Nima Sarshar and Xiaolin Wu \\
Department of Electrical and Computer Engineering \\
McMaster University, Hamilton, Ontario, Canada \\
{\small
nima@grads.ece.mcmaster.ca/xwu@mail.ece.mcmaster.ca}\vspace*{-.2in}
}

\maketitle

\begin{abstract}
We are interested in how to best communicate a (usually real valued)
source to a number of destinations (sinks) over a network with
capacity constraints in a collective fidelity metric over all the
sinks, a problem which we call joint network-source coding. Unlike
the lossless network coding problem, lossy reconstruction of the
source at the sinks is permitted.  We make a first attempt to
characterize the set of all distortions achievable by a set of sinks
in a given network.  While the entire region of all achievable
distortions remains largely an open problem, we find a large,
non-trivial subset of it using ideas in multiple description coding.
The achievable region is derived over all balanced
multiple-description codes and over all network flows, while the
network nodes are allowed to forward and duplicate data packets.

\end{abstract}
\section{Introduction}
\subsection{Joint Network-Source Coding: Problem
Formulation}\label{sec:form}

Joint network-source coding (JNSC) is the problem of communicating
and reconstructing a (usually real valued) source in a network to
a maximal collective fidelity over a given set of sinks, while the
flows of the code streams satisfy the edge capacities of the
network.  JNSC can be considered as a lossy version of the
(lossless) network coding problem, since the reconstruction is not
necessarily perfect. The source is "observed" by a subset of nodes
in the network, called source nodes. Due to capacity constraints,
source nodes have to communicate a coded version of the source to
their neighboring nodes.  Just as in lossless networked coding,
intermediate nodes can in general transcode data received from
other nodes, and communicate it to their neighbors. Any node in
the network, based on the information it receives about the
source, can reconstruct the source with some distortion. This
paper aims to characterize the set of distortions simultaneously
achievable at a pre-specified subset of nodes (sinks) in the
network.

Unlike its lossless counterpart, the interaction of lossy
source-network codes with arbitrary networks is largely
unexplored.
In fact, the term network coding refers, almost exclusively, to
lossless network communication. This is despite the fact that
arguably the majority of
applications, both in the Internet and in various wireless setups,
involve lossy source communication, in particular for multimedia
applications. It should however be noted that some of the well
studied examples of multi-terminal   source coding problems (e.g.,
multiple-description coding) are simple examples of a general
lossy networked coding problem.

In this paper the network model is similar to, now standard,
models in network coding \cite{ash}.  The JNSC problem is defined
by the following elements: \\ (1) A directed graph $G\langle V,E
\rangle$.\\ (2) A function $R:E\rightarrow\mathbb{R}^+$ that
assigns a capacity $R(e)$ to each link $e\in E$. We normalize
bandwidth with the source bandwidth, therefore, $R(e)$ is
expressed in units of bits per source symbol. \\ (3) A source $X$
in some alphabet $\Gamma$ and a set of distortion measures
$\rho^n:\Gamma^n \rightarrow \mathbb{R}^+$. We assume $X$ admits a
rate-distortion function $D_X(R)$, with $\rho$ as the measure. \\
(4) Two sets $S,T \subseteq V$ that denote the set of source and
sink nodes respectively.  The source nodes observe, encode, and
communicate $X$ in the network. Source nodes are assumed to be
able to collaborate in encoding. This can model, for example,
computer networks where sources are encoded off-line and copies of
the code are distributed to the source nodes.

Nodes can communicate with neighbor nodes at a rate specified by
the capacity of the corresponding link. The goal is to communicate
the source $X$ from the source nodes in $S$, and reconstruct $X$
at the sink nodes in $T$. A distortion vector
$\mathbf{d}=(d_t,t\in T)\in \mathbb{R}^{|T|}$ is said to be
achievable if $X$ can be reconstructed with a maximum distortion
of $d_t$ at a sink node $t$ by using a coding scheme that respects
the capacity constraints on the links, i.e., the rate of
information per source sample communicated over $e$ is less than
$R(e)$.  As in \cite{ash}, we need to leave the details of the
code unspecified, because it proves extremely hard to come up with
the most general class of possible codes. An intriguing problem is
how to characterize the set of all achievable distortion
$t$-tuples ${\cal D}_X(G,S,T,R) \subset \mathbb{R}^{|T|}$.   Note
that this problem includes the usual lossless network coding
problem if the source alphabet $\Gamma$ is finite and the
distortion measure is defined so that $\rho^n(A^n,B^n)=0$ if and
only if $A^n=B^n$.

\subsection{Multiple-Descriptions: a Tool for JNSC}

Multiple-description codes (MDC) have always been associated with
robust networked communications, because they are designed to
exploit the path and server diversities of a network.  The present
active research on MDC is driven by growing demands for real-time
multimedia communications over packet-switched lossy networks,
like the Internet.  With MDC, a source signal is encoded into a
number of code streams called descriptions, and transmitted from
one or more source nodes to one or more destinations in a network.
An approximation to the source can be reconstructed from any
subset of these descriptions.
If some of the descriptions are lost,
the source can still be approximated by those received. This is
why there seems to be a form of consensus in the literature in
that multiple description codes should only be used in
applications involving packet loss, because only in this case the
overhead in the communication volume can be justified.

\begin{figure}
\begin{center}
\includegraphics[width=2.5in,height=1.5in]{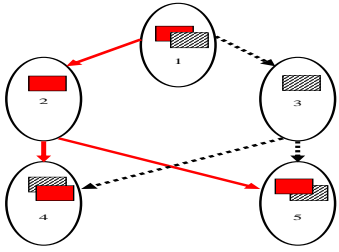}
\caption{An example of flow of a two description
code.}\label{fig:1}
\end{center}\vspace{-.3in}
\end{figure}

This paper shows, however, that MDC is beneficial for lossy
communication even in networks where all communication links are
error free with no packet loss.  In this case, multiple
description coding, aided by optimized routing, can improve the
overall rate-distortion performance by exploiting various paths to
different nodes in the network.  This can be easily demonstrated
through an example. In Fig.\ \ref{fig:1}, a source node (node 1)
feeds a coded source into a network of four sink nodes (nodes
2-5). The goal is to have the best reconstruction of the source at
each of these four nodes.  All link capacities are $C$ bits per
source symbol. MDC encodes the source into two descriptions (shown
by solid and dashed boxes in the figure), each of rate $C$.
Descriptions 1 and 2 are sent to nodes 2 and 3 respectively. Node
2 in turn sends a copy of description 1 to nodes 4 and 5, while
node 3 also sends a copy of description 2 to nodes 4 and 5.  In
the end, nodes 4 and 5 will each receive both descriptions, while
nodes 2 and 3 will only receive one description.

To see how the nodes in the network benefit from MDC, let
$D_1(C),D_2(C),D_{12}(C)$ be the distortion in reconstructing the
source given description 1 or 2 or both.  Let
$\mathbf{d}=(d_2,d_3,d_4,d_5)$ be the vector of the average
distortions in reconstructing the source at nodes 2 through 5.
Therefore, $\mathbf{d}=(D_1(C),D_2(C),D_{12}(C),D_{12}(C))$.

Let's define $\mathcal{D}_M$ as the set of all achievable
distortion 4-tuples $\mathbf{d}$.  Although MDC is in general a
special form of lossy networked coding, $\mathcal{D}_M$ still
contains a large and interesting subset of all achievable
distortion tuples. In this example, it includes for instance, the
distortion region achievable by separate source and networked
 coding. By results in \cite{ash}, the maximum rate
with which common information can be communicated to nodes 2
through 5 is $C$ bits per source symbol. Therefore, the distortion
rate achievable by separate source and network coding is
$\mathcal{D}_S =\{(\delta_2,\delta_2,\delta_4,\delta_5):
\delta_i\geq 2^{-2C}, i=2,3,4,5\}$.  We immediately have that
$\mathcal{D}_S \subset \mathcal{D}_M$ by noting that
$D_1(C)=D_2(C)=D_{12}(C)=2^{-2C}$ is part of $\mathcal{D}_M$. In
fact this corresponds to communicating two identical descriptions,
each of which is an optimal (in the rate-distortion sense) source
code of rate $C$ for $X$.

The inefficiency of separate source and network coding lies in
that even through nodes $4,5$ have twice the incoming capacity
compared to nodes $2,3$, their reconstruction error ($d_4=d_5$) is
bounded by the reconstruction error of the weaker nodes
($d_2=d_3$).  Unlike lossless coding, lossy codes can play a
tradeoff between the reconstruction errors at different nodes,
generating a much larger set of achievable distortion tuples
$\mathbf{d}$ than $\mathcal{D}_S$. These tradeoffs are essential
in practice. For instance, in networked multimedia applications
over the Internet, where the network consists of a set of
heterogenous nodes, the experience of a user with broadband
connection should not be bounded by that of a user with a lesser
bandwidth. Such tradeoffs are perhaps best treated as an
optimization problem by introducing appropriate Lagrangian
multipliers (or weighting functions). An objective function to
minimize, therefore, can be defined as
\begin{equation}\label{eqn:ovd}
\overline{d}(\mathbf{p},\mathbf{d})=\mathbf{p}^T \cdot \mathbf{d}
\end{equation}
where $\mathbf{p}=[p_2,p_3,p_4,p_5]$ is an appropriate weighting
vector. An optimal solution will be given by:
\[
\mathbf{d}^*(\mathbf{p})=\arg\min_{\mathbf{d} \in \mathcal{D}_M}
\overline{d}(\mathbf{p}, \mathbf{d})
\]

Once the optimal distortion vector $\mathbf{d}^*$ is found, one
should, in principle, be able to find a multiple description code
that provides the marginal and joint distortions corresponding to
$\mathbf{d}^*(\mathbf{p})$ (such an MDC exists).

As a concrete example, let's optimize the average distortion at
all nodes 2 through 5 in Fig. \ref{fig:1} for
$\mathbf{p}=[1/4,1/4,1/4,1/4]$, in which case:
\begin{equation}\label{eqn:1}
\overline{d}=\frac{2D_{12}(C)+D_1(C)+D_2(C)}{4}
\end{equation}
To be specific, lets assume that the source in question is an iid
Gaussian with variance one for which achiveable distortions in
multiple description coding are completely derived by Ozarow in
\cite{oz}. The symmetry in indices 1 and 2 ensures that
(\ref{eqn:1}) is minimized when the two descriptions are balanced,
that is, $D_1(C)=D_2(C)=D$.  Ozarow's result, when specialized to
balanced MDC states that the following set of distortions are
achievable:
\begin{eqnarray}\label{eqn:2}
D_1 &=& D_2 = D \geq 2^{-2C} \\
D_{12} &\geq&
\frac{2^{-4C}}{(D+\sqrt{D^2-2^{-4C}})(2-D-\sqrt{D^2-2^{-4C}})}\nonumber
\end{eqnarray}

The average distortion in (\ref{eqn:1}) can therefore be minimized
under the constraints of (\ref{eqn:2}). This is a particularly
easy task because the region (\ref{eqn:2}) is convex. Let this
optimal average distortion be $\overline{d}_M^*(C)$. By separating
source from network coding, the reconstruction distortion at nodes
2 through 5 (and hence the average distortion over all these
nodes) is at best $d_S(C)=2^{-2C}$.  It is easy to show that
$\overline{d}_M^*(C) < d_S(C)$ for all $C>0$.  In other words, for
all $C>0$ there exists a balanced two description code for which
the average distortion over all sink nodes is strictly less than
the average distortion achievable by any separate source and
network  coding scheme.

A number of important observations are due:

$\centerdot$ MDC routing can exploit path diversity in ways that a
separate source and network coding can not. For instance, in the
example of Fig. (\ref{fig:1}), nodes 4 and 5 can benefit from the
data received both from nodes 2 and 3, while nodes 2 and 3
themselves can benefit from the data they relay, which was not
possible if a common data was communicated to both nodes 2 and 3
by the source
node.\\
$\centerdot$ To benefit from MDC in the network, routing needs to
be optimized.\\ $\centerdot$ Not only the routing, but also the
MDC should be designed optimally. In our example, we did this by
choosing an MDC with desired side and joint distortions to
minimize the average distortion (\ref{eqn:1}). In other words,
while any distortion pair satisfying (\ref{eqn:2}) is achievable
by some MDC, only one pair $(D,D_{12})$ (corresponding to a
particular MDC design) can minimize (\ref{eqn:2}).\\
$\centerdot$ Unlike the case of minimizing (\ref{eqn:2}),
optimizing the MDC may result in a different number of,
potentially unbalanced, descriptions. In general, the total number
of descriptions and their rates are left as optimization
parameters.

In this paper, by confining ourselves to balanced MDC codes of the
same rate, we will be able to find a practically interesting,
achievable distortion region for the JNSC problem.
In Section (\ref{sec:rnf}) we introduce both discrete and
continuous versions of a new routing problem, called rainbow
network flow, which plays an integral role in JNSC.  We then state
our main achievability results which are proved in Section
(\ref{sec:ach}).

\section{Rainbow Network Flow Problem}\label{sec:rnf}
The ideas in the previous section are formalized into the concept
of  Rainbow Network Flow (RNF). RNF is concerned with routing and
duplication of balanced MDC description packets in an arbitrary
network and the subset of descriptions received by sink nodes. RNF
for balanced descriptions of the same rate is
posed in the following setting:\\
(1) $G\langle V,E\rangle $, a directed graph with a node set $V$
and an edge set $E$.\\
(2) $S=\{s_1,s_2,...,s_{|S|}\},T=\{t_1,t_2,t_3,...,t_{|T|}\}$ two
subsets of $V$ representing the set of source and sink nodes
respectively.\\ (3)  A function $R:E\rightarrow \mathbb{R}^+$
representing the capacity of each link in $G$.\\ (4) A set
$\chi\subset \mathbb{R}$ called the description set.\\ (5) An
$r\in \mathbb{R}^+$ called the description rate.
\\ (6) $\mu:{\cal P}(\chi)\rightarrow \mathbb{R}^+$, a measure on
$\chi$, where ${\cal P}(\chi)$ denotes the set of all subsets of
$\chi$.\\

A flow path from $s\in S$ to $t\in T$ is a sequence of edges
$w(s,t)=[(v_0=s,v_1),(v_1,v_2),...,(v_{m-1},v_{m}=t)]$, such that
$(v_i,v_{i+1})\in E$ for $i=0,1,...,m-1$.



A rainbow network flow (RNF), denoted by
$\alpha(G,T,S,\chi,r,W,f)$, consists of a set $W$ of flow paths in
$G$,
and a so-called flow coloring function $f: W \rightarrow \chi$.
For the RNF in Fig.\ \ref{fig:1}, $W=\{ [(1,2),(2,4)]$,
$[(1,2),(2,5)]$, $[(1,3),(3,4)]$, $[(1,3),(3,5)] \}$, and
the flow coloring function $f$ assigns $f([(1,2),(2,4)])=1$,
$f([(1,2),(2,5)])=1$, $f([(1,3),(3,4)])=2$, $f([(1,3),(3,5)])=2$.

The rainbow network flow problem is said to be discrete (dRNF) if
$\chi\subset \mathbb{N}$ and $\mu({\cal M})=r|{\cal M}|$ where
$|\cdot|$ is the cardinality of a finite set. The example in Fig.\
\ref{fig:1} corresponds to a dRNF with $\chi=\{1,2\}$, $r=C$,
$S=\{1\}$, $T=\{2,3,4,5\}$, $R(e)=C$ for all $e\in E$.

The rainbow network flow problem is said to be continuous (cRNF)
if $\chi$ is the Borel algebra on $R$ and $\mu=\mu_B$ is the Borel
measure. The parameter $r$ becomes irrelevant in cRNF.

Throughout the paper, we will liberally drop the arguments when
they are obvious from the context or are not relevant to the
formulation at hand.


Let $\Phi_E(e,W)$ and $\Phi_V(v,W)$ be the sets of all colored
flow paths in $W$ that contain the link $e$ or the node $v$,
respectively.  For example,
$\Phi_E(e=(1,2),W)=\{[(1,2),(2,3)],[(1,2),(2,5)]\}$.

The spectrum of an edge $e\in E$, with respect to RNF $\alpha$, is
defined as:
\[
\Psi_E(\alpha,e) \equiv \bigcup_{w \in \Phi_E (e,W)} f(w)
\]
Likewise, the spectrum of a node $v$ is defined as:
\[
\Psi_V(\alpha,v) \equiv \bigcup_{w\in \Phi_V (v,W)} f(w)
\]

In Fig.\ \ref{fig:1} for instance
$\Psi_E((1,2))=\Psi_E((2,4))=\Psi_E((2,5))=\{1\}$ and,
$\Psi_E((1,3))=\Psi_E((3,4))=\Psi_E((3,5))=\{2\}$.  The spectrum
of the nodes $4,5$ consists of both descriptions (i.e.,
$\{1,2\}$), while the spectrum of the nodes $2,3$ is $\{1\},\{2\}$
respectively.

An RNF $\alpha(G,R,W,f)$ is said to be admissible with capacity
function $R$, if and only if:
\begin{equation}\label{eqn:cap}
\mu(\Psi_E(\alpha,e))<R(e) \qquad \forall e\in E
\end{equation}

The significance of this inequality is that it allows for
duplication of a description by relay nodes. Therefore, two flow
paths of the same color can pass through a link $e$, and yet
consume a bandwidth of only $r$.

The RNF plotted in Fig.~\ref{fig:1} is admissible because at most
one description with rate $C$ is communicated over each link and
the capacity of each link is $C$.  This is made possible by
duplicating at nodes 2,3.

\begin{figure}
\begin{center}
\includegraphics[width=2.75in,height=3.8in]{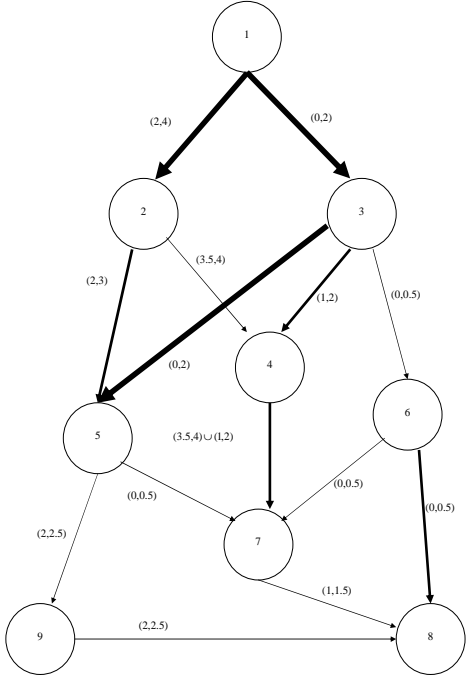}
\caption{Example of an admissible cRNF: The capacity of the edges
are 2,1,0.5 which is proportional to the thickness of the line.
Node $1$ is a source and nodes 6,7,8 are sinks. On each edge, the
subset of the real line that is routed over that edge is shown.
}\label{fig:cont}
\end{center}\vspace{-.3in}
\end{figure}

Let ${\cal F}(G,R)$ be the set of all admissible RNF's in $G$ with
capacity function $R$.  Any RNF $\alpha\in {\cal F}(G,R)$ results
in an admissible rainbow flow vector (RFV),
$\mathbf{q}(\alpha)=(q_{t};t\in T) \in \mathbb{R}^{|T|}$, such
that:
\begin{equation}\label{eqn:rfv}
q_t= \mu (\Psi_V(\alpha,t))
\end{equation}
In Fig.\ \ref{fig:1} for instance, the depicted flow results in an
RFV $\mathbf{q}=C(1,1,2,2)$ (i.e., nodes 2,3 receive one
description and nodes 4,5 receive two).  Fig.\ \ref{fig:cont}
depicts an example of an admissible cRNF.  The spectrum of the
nodes $6,7,8$ are: $\Psi_V(v_6)=(0.5,2) \cup (3,4)$,
$\Psi_V(v_7)=(2,2.5)$, and $\Psi_V(v_8)=(1,1.5) \cup (2,2.5)$
which results in an RFV of $\mathbf{q}=(1.5,0.5,2.5)$.

Let ${\cal Q}(G,R) \subset \mathbb{R}^{|T|}$ be the set of all
admissible RFV's:
\[
{\cal Q}(G,R) \equiv \bigcup_{\alpha\in {\cal F}(G,R)}
\mathbf{q}(\alpha)
\]

\section{Achievability Results}\label{sec:ach}

To proceed we need a way of parameterizing the family of MDC's
that are generated by the well-known technique of Priority
Encoding Transmission (PET).


Let ${\cal H}$ be the set of all functions
$y:\mathbb{\chi}\rightarrow \mathbb{R}^+$, such that $\int_{
\chi}yd\mu=1$.  For dRNF, in particular, the set ${\cal
H}=\dot{{\cal H}}$ is the set of all discrete vectors
$y=(y_i;i=1,2,...,|\chi|)$ such that $\sum_{i=1}^{|\chi|} y_i=1$.
For cRNF, ${\cal H}$ consists of all functions $y$ such that
$\int_{0}^\infty y(r)dr=1$.

We can show that there is a one to one correspondence between the
members of set $\dot{{\cal H}}$ and the multiple description codes
generated by the PET technique. Furthermore, the reconstruction
distortion at each sink node can be calculated given the RFV
$\mathbf{q}$ and a function $y\in\dot{{\cal H}}$.

For any $y \in {\cal H}$, $q\in {\cal Q}(G,R)$, define the
$|T|$-tuple real valued vector $\mathbf{d}=(d_t;t\in T)$ such
that:
\[
d_t(y,\mathbf{q})=D_X\left(\int_{x\in \chi:\mu(x)<q_t}\mu(x) y(x)
d\mu \right)
\]
where $D_X(\cdot)$ is the distortion-rate function of the source
$X$.

For dRNF, $\mathbf{d}=\dot{{\mathbf d}}(y,\mathbf{q})=(\dot{d}_t;
t\in T) \in \mathbb{R}^{|T|}$ such that:
\begin{equation}\label{eqn:del}
\dot{d}_t(y,\mathbf{q}) \equiv D_X \left(r
\sum_{i=0}^{q_{t}/r}iy(i)\right)
\end{equation}

Note that for dRNF $q_t/r$ is an integer equal to the number of
distinct descriptions node $t$ received. Let $\dot{{\cal
B}}(G,r,\chi)\subset \mathbb{R}^{|T|}$ be the union of all such
vectors $\dot{\mathbf{d}}$, that is,
\[
\dot{{\cal B}}(G,r,\chi)\triangleq \bigcup_{y\in \dot{{\cal H}},q
\in {\cal Q}(G,R)}\dot{\mathbf{d}}(y,\mathbf{q}).
\]

It should be evident by now that for dRNF, $\dot{{\cal
B}}(G,r,|\chi|)$ depends only on the cardinality of the set $\chi$
and not the actual set of descriptions.

Likewise for cRNF, the vector $\mathbf{d}=
\mathbf{d}(y,\mathbf{q})=(d_t; t\in T)$ is defined such that:
\begin{equation}\label{eqn:delc}
d_t(y,\mathbf{q}) \equiv D_X\left(\int_{x=0}^{q_{t}}xy(x)dx\right)
\end{equation}
Also, we define ${\cal B}(G) \subset \mathbb{R}^{|T|}$ to be the
union of all $\mathbf{d}(y,\mathbf{q})$ over $y \in {\cal H}$ and
$\mathbf{q} \in {\cal Q}(G,R)$.

The following is our main theorem.\\
\begin{thm}\label{main}
${\cal B}(G) \subset {\cal D}_X(G,S,T,R)$. \\
\end{thm}


Here, ${\cal B}(G)$ represents a large class of achievable
distortion vectors for the JNSC problem. To prove theorem
\ref{main}, we need to establish a series of theorems and lemmas.
The first step is to establish the achievability result for the
discrete version of the problem. \\

\begin{thm}\label{thm:dis}
 $\dot{{\cal B}}(G,r,|\chi|)\subset {\cal D}_X(G,R,S,T)$ for
  all $r\in \mathbb{R}^+$ and
 $\chi\subset \mathbb{N}$.\\
\end{thm}
\begin{proof}
The achievability is a direct consequence of our construction. In
particular  any achievable RFV $\mathbf{q}=(q_t,i\in T)$, defined
in (\ref{eqn:rfv}), indicates that $q_t/r$ distinct descriptions
each of rate $r$ can be communicated to sink nodes $t\in T$. To
find the reconstruction distortion at a node $t\in T$ using this
MDC, we would need to know the exact subset of $\chi$ that is
present at $t$. However, $q_t/r$  only specifies the total
\textit{number} of the distinct descriptions received and not the
exact subset of the descriptions. That is why we need to assume
that the MDC is \textbf{\textit{balanced}}, that is, for any
$0\leq k\leq |\chi|$, the source can be reconstructed to the same
fidelity given any subset of size $k$ out of the total of $|\chi|$
descriptions. In this case, knowing the total number of
\emph{distinct} descriptions available at $t\in T$ (i.e., $q_t/r$)
suffices to find the reconstruction distortion at $t$.

For producing balanced MDC, we use a popular method called
Priority Encoding Transmission (PET), with which any number of
balanced multiple descriptions can be produced from a
progressively encoded source stream. The idea is the following. To
make $|\chi|$ balanced descriptions each of rate $r$ bits per
source symbol, for a large enough value of $n$, encode $n$ samples
of $X$ into a progressive bitstream $(b_0,b_1,...,b_{nrL})$, where
we have assumed $n\cdot r$ is an integer for simplicity. Let
$K=|\chi|$. Then take an $K\cdot n\cdot r$ binary matrix and call
it $Y=[Y_{ij},i=1,2,...,K,j=1,2,...,n\cdot r]$. Now take
$y=(y_i,i=1,2,...,K)$, any vector of real numbers of length $K$
such that $\sum_{i=1}^K y_i=1$. Now, for $i=1,2,...,K$ do the
following: let $Y_l,Y'_l$ for $l=1,2,...,K$ be sub-matrices of $Y$
consisting of: \begin{eqnarray*}Y_l=[Y_{ij}&;&\: i=1:l,\quad
j=\sum_{k=1}^{l-1} n\cdot r y_k:\sum_{k=1}^{l} n\cdot r y_k]\\
Y'_l=[Y'_{ij}&;&\: i=l+1:K,\quad j=\sum_{k=1}^{l-1} n\cdot r
y_k:\sum_{k=1}^{l} n\cdot r y_k]
\end{eqnarray*}

Therefore, matrix $Y_i$ will contain $i\times n \cdot r\times y_i$
bits while $Y'_i$ has $(K-i)\times n\cdot r\times y_i$ bits. For
$i=1,2,...,K$, put the $i\times n \cdot r\times y_i$ bits of the
progressive source code stream , from $b_{g(i)}$ to
$b_{g(i)+i\times n\cdot r\times y_i}$ in $Y_i$, where
$g(i)=\sum_{k=1}^i k\times n\cdot r y_k$. In $Y'_i$ on the other
hand, put parity symbols of a $(i\cdot n\cdot r\cdot y_i,K\cdot
n\cdot r\cdot y_i )$ ideal \textit{erasure correction} code
corresponding to the bits in $Y_i$.

Now the descriptions consist of the $K$ columns of the matrix $Y$,
each of $n\cdot r$ bits. The total source bits used is $n\cdot r
\sum_{k=1}^K k y_k$. It is easily verified that given any $l\leq
K$ descriptions, the first $\xi_l=\sum_{k=1}^l k\cdot n\cdot
r\cdot y_k$ bits of the source bitstream can be recovered. For
large enough $n$ and assuming the source is progressively
refinable, given any $k$ distinct descriptions, the source can
therefore be reconstructed within  distortion:
\begin{equation}\label{eqn:dx} D_X(\xi_k/n)=D_X\left(r \sum_{l=1}^k
l y_l\right)\end{equation}

We denote this choice of MDC by ${\cal U}(y,r,|\chi|,X)$.  Any
non-negative vector $\mathbf{y}$, such that $\sum_{i=1}^{|\chi|}
y_i=1$ therefore specifies a valid ${\cal U}(y,r,K,X)$ code and vice
versa. Theorem \ref{thm:dis} is now easily proved by comparing
(\ref{eqn:dx}) and (\ref{eqn:del}).
\end{proof}

Then we show that the achievable region for dRNF will converge to
that for cRNF as stated by the following theorem.\\

\begin{thm}\label{thm:conv}
$\forall r>0$ and $\chi\subset \mathbb{N}$,\[\dot{\cal
B}(G,r,|\chi|)\subset\dot{\cal B}(G,r,\infty)\subset {\cal B}(G)
\] and moreover,
$\lim_{r\rightarrow 0^+}\dot{\cal B}(G,r,\lfloor g /r \rfloor)
={\cal B}(G)$
 for some constant $g$ depending on $G,R$ only.\\
\end{thm}

Theorem \ref{main} follows directly from Theorems \ref{thm:conv}
and \ref{thm:dis}. The proof of Theorem \ref{thm:conv} is
essentially based on two facts, (1) any cRNF can be approximated
arbitrarily closely by a dRNF with small enough packet size $r$
and, (2) reducing the packet size $r$ dose not decrease the
achievable region of dRNF. In the remaining of this paper, we will
put forward the main lemmas required for proving Theorem
\ref{thm:conv}. Unfortunately, due to lack of space, the details
of some of these lemmas have to be published elsewhere.

 First, we need to show that
increasing the number of descriptions can only increase the
achievable region. Therefore, in dRNF, there is no harm in letting
the set $\chi$ to be the whole integers $\mathbb{N}$.

\begin{lem}\label{lem:pac} For all $r\in \mathbb{R}^+$ if $K>K'\in \mathbb{N}$
then: $\dot{\cal B}(G,r,K')\subset {\cal B}(G,r,K)\subset\dot{\cal
B}(G,r,\infty)$.
\end{lem}
\begin{proof} Take $\mathbf{d}$ any member of ${\cal
B}(G,r,K)$ which corresponds to some ${\cal U}(y,r,K,X)$ and a flow
vector $\mathbf{q}\in{\cal Q}(G,R)$. Trivially, for $K'>K$, the flow
vector $\mathbf{q}$ remains achievable. Define ${\cal U}(y',r,K',X)$
such that $y'_{l}=y_{l},l=1,2,...,K$ and $y'_{l}=0, l=K+1,...,K'$.
The distortion vector $\mathbf{\delta'}\in {\cal B}^{D}(G,r,X,K') $
corresponds to ${\cal U}(y',r,K',X)$ and the flow vector
$\mathbf{q}'$ is therefore equal to $\mathbf{q}$, proving the
assertion.
\end{proof}

Next, we show that by dividing the rate of the descriptions $r$ by
an integer number $i$, and multiplying the total number of
descriptions by $i$, the set of achievable distortions can only
increase.

\begin{lem}\label{lem:div} For any integers $K,i\in \mathbb{N}$ and real number $r>0$, one has
$\dot{\cal B}(G,r,K)\subset \dot{\cal B}(G,r/i,i.K)\subset\dot{\cal
B}(G,r/i,\infty)$.
\end{lem}

\begin{proof} Note that any flow path $w$ can be split into $i$ flows, each of rate $r/i$. Any
admissible flow therefore, will still remain admissible. Now,
expand the set $\chi$ to a set $\chi'$ such that
$|\chi'|=i|\chi|$. To each member of $\chi$, we can therefore
assign $i$ distinct members of $\chi'$. For any admissible flow
vector therefore, $\mathbf{q}\in{\cal Q}(G,R)$ and for any integer
$i$, there exists an admissible flow vector $\mathbf{q'}\in{\cal
Q}(G,r/i,iL)$ such that $q'_t=q_t$. Now let $\mathbf{d}$ be a
distortion vector corresponding to $\mathbf{q}$ and ${\cal
U}(y,r,K,X)$. Then define ${\cal U}'(y',r/i,iL,X)$, such that
$y'_{ik}=y_{k}$ for all $k=1,2,...,K$ and we let $y'_{k'}=0$ for
all $k'$ not divisible by $i$. The distortion vector $\mathbf{d}'$
corresponding to ${\cal U}'(y',r/i,iK,X)$ and $\mathbf{q}$ is
therefore equal to $\mathbf{\delta}$. Therefore, any distortion
vector $\mathbf{\delta}\in \dot{\cal B}(G,r,K)$ is equal to a
distortion vector $\mathbf{d}'$ in $\dot{\cal B}(G,r/i,X,i.K)$
which proves the assertion.

The last part follows from Lemma (\ref{lem:pac}), that is,
allowing for more number of descriptions will not decrease the
achievable distortion region.
\end{proof}

 For any $r$, therefore,
the largest achievable region will occur when $i$ goes to
infinity, or the size of the descriptions $r/i$ goes to zero. This
however, does not necessarily mean that the limit
$\lim_{r\rightarrow 0^+}\dot{\cal B}(G,r,\infty)$ exists. In fact,
$\dot{\cal B}(G,r,K)$ is not necessarily continuous for all $r>0$.
Take for example the extreme case where the size of the
description $r$ is equal to the maximum outgoing capacity of all
source nodes. Then, no description packet is able to leave any
source. When $r$ is slightly decreased however, it might be
possible to have a single description flow and therefore the set
of achievable distortions can jump discontinuously. As the
description size becomes smaller, however, this discontinuity
becomes less and less significant. The following lemma formalizes
this.
\begin{lem}\label{lem:convg} Take any $r'<r$ and a $\mathbf{d}'\in \dot{\cal
B}(G,r',\infty)$. Then there exists a $\mathbf{d}\in \dot{\cal
B}(G,r,\infty)$ such that $\forall t\in T,|d_t-d'_t|\leq
\gamma\cdot r$ for some constant $\gamma>0$ depending only on the
network topology.
\end{lem}
\begin{proof}
Let $\mathbf{d}'\in \dot{\cal B}(G,r',\infty)$ correspond to a
flow vector $\mathbf{q}'$ and an MDC ${\cal U}(y,r',\infty,X)$.
Now if the size of the descriptions where increased from $r'$ to
$r>r'$, the measure of the spectrum at every edge $e$ (see
\ref{eqn:rfv}) should be multiplied by at most
$|\Psi_E(\alpha,e)|\times (r/r')$. This of course might render the
flow infeasible because some of the conditions in
(\label{eqn:rfv}) are not satisfied any more. For any such
overloaded edge, we delete some of the flows randomly until the
capacity constraint on the edge is respected. 

We know that a total spectrum of $|\Psi_E(\alpha,e)|$ using
descriptions of rate $r'$ can flow through $e$. When we replace
these descriptions with descriptions of rate $r>r'$, we can keep
at least $\lfloor |\Psi_E(\alpha,e)/r| \rfloor$ of these flow
paths, that correspond to \emph{distinct} descriptions. The total
spectrum of $e$ therefore is at least
$(|\Psi_E(\alpha,e)|/r-1)\times r=|\Psi(\alpha,e)|-r$. In other
words, there is a flow with descriptions of rate $r$ that respects
the capacity on the edges $e$ and does not decrease the total
spectrum of $e$ by more than $r$.

Now replacing flows of rate $r'$ with flows of rate $r>r'$ for edge
$e$ might also disrespect the constraint on other edges. For each
such edge, however, we can repeat the procedure, that is replacing
all the flows with rate $r'$ with flows of rate $r$ and then
deleting extra flows until the capacity on that edge is respected.
Since there are at most $|E|$ edges, after replacing all the flows
with flows of rate $r$ and deleting excess flows, the total
reduction in the overall flow is at most $r\times |E|^2$.

Now for any ${\cal U}(y,r',\infty,X)$, a distortion vector
$\mathbf{d}'(\delta'_t;t\in T)$ is obtained. Replacing the old
flow with this new flow, a new distortion vector
$\mathbf{d}=[\delta_t; t\in T]$ can be achieved such that
\begin{eqnarray*}
d_t&\leq &D_X\left(r'\sum_{k=1}^{q_t/r}k
y_i-r|E|^2\right)\\
&\leq& D_X\left(r\sum_{k=1}^{q_t/r}k y_i\right)+Z_X |E|^2r+o(r)\\
&=& d'_t+\gamma\times r+o(r)
\end{eqnarray*}
where in the last inequality we have used the fact that $D_X(.)$
is differentiable with a bounded derivative and $Z_X$ is a
constant depending on the function $D_X$ and $\gamma=Z_X|E|^2$ is
independent of $r$.
\end{proof}
The following lemma can therefore be proved:
\begin{lem}\label{lem:all}
Take any countable sequence $r_1,r_2,...$ of positive numbers,
 such that $\lim_{n\rightarrow \infty} r_n=0$. Then,
 $
\bigcup_n \dot{\cal
 B}(G,r_n,\lfloor |E|^2/r_n \rfloor)
 $
 converges to ${\cal B}(G)$.
 \end{lem}
 \begin{proof} Using (\ref{lem:convg}), one can show that $\bigcup_n \dot{\cal
 B}(G,r_n,\lfloor |E|^2/r_n \rfloor)$ converges to a unique ${\cal B}^*(G)$ regardless of the sequence $r_n$. Now note that the cRNF routes subsets of a Borel $\sigma$-algebra.
 As such, each flow in cRNF with finite Borel measure can be
approximated arbitrarily closely with a collection of discrete
flows, each of rate $r_n$ provided that $\lim_{n\rightarrow \infty}
r_n=0$. One can use this to show that the limit ${\cal B}^*(G)$ is
in fact equal to ${\cal B}(G)$ of cRNF. The details of the proof are
omitted due to space limit.
\end{proof}
Theorem \ref{thm:conv} is a direct consequence of Lemmas
\ref{lem:all}. Together, Theorems \ref{thm:dis} and \ref{thm:conv}
lead to Theorem \ref{main}.

\vspace{-.1in}
\bibliographystyle{abbrv}

\end{document}